\documentclass[a4paper,twoside]{article}

\usepackage{epsfig}
\usepackage{subcaption}
\usepackage{calc}
\usepackage{amssymb}
\usepackage{amstext}
\usepackage{amsmath}
\usepackage{amsthm}
\usepackage{multicol}
\usepackage{pslatex}
\usepackage{apalike}
\usepackage{algorithm2e}
\usepackage[bottom]{footmisc}
\usepackage{SCITEPRESS}     

\usepackage{graphicx}
\graphicspath{ {./img/} }
\usepackage{float}
\usepackage{stfloats}
\usepackage{tabularx,booktabs}
\newcolumntype{Y}{>{\centering\arraybackslash}X}

\usepackage{calrsfs}
\DeclareMathOperator*{\argmax}{argmax}

\begin{document}

\title{A machine learning workflow to address credit default prediction}

\author{\authorname{Rambod Rahmani\sup{1}\orcidAuthor{0009-0009-2789-5397}, Marco Parola\sup{1}\orcidAuthor{0000-0003-4871-4902} and Mario G.C.A. Cimino\sup{1}\orcidAuthor{0000-0002-1031-1959}}
\affiliation{\sup{1}Dept. of Information Engineering, University of Pisa, Largo L. Lazzarino 1, Pisa, Italy}
\email{\{r.rahmani@studenti, marco.parola@.ing, mario.cimino@\}.unipi.it}
}


\abstract{Due to the recent increase in interest in Financial Technology (FinTech), applications like credit default prediction (CDP) are gaining significant industrial and academic attention.
In this regard, CDP plays a crucial role in assessing the creditworthiness of individuals and businesses, enabling lenders to make informed decisions regarding loan approvals and risk management.
In this paper, we propose a workflow-based approach to improve CDP, which refers to the task of assessing the probability that a borrower will default on his or her credit obligations. 
The workflow consists of multiple steps, each designed to leverage the strengths of different techniques featured in machine learning pipelines and, thus best solve the CDP task.
We employ a comprehensive and systematic approach starting with data preprocessing using Weight of Evidence encoding, a technique that ensures in a single-shot data scaling by removing outliers, handling missing values, and making data uniform for models working with different data types.
Next, we train several families of learning models, introducing ensemble techniques to build more robust models and hyperparameter optimization via multi-objective genetic algorithms to consider both predictive accuracy and financial aspects.
Our research aims at contributing to the FinTech industry in providing a tool to move toward more accurate and reliable credit risk assessment, benefiting both lenders and borrowers.}

\onecolumn \maketitle \normalsize \setcounter{footnote}{0} \vfill


\section{Introduction and background}\label{sec:intro}

In the financial sector, credit scoring is a crucial task in which lenders must assess the creditworthiness of potential borrowers.
In order to determine credit risk, several characteristics related to income, credit history, and other relevant aspects of the borrower must be deeply investigated.

To manage financial risks and make critical decisions about whether to lend money to their customers, banks and other financial organizations must gather consumer information to identify reliable borrowers from those unable to repay debt. This results in solving a credit default prediction problem, or in other words a binary classification problem \cite{moula2017credit}.

In order to address this challenge, over the years several statistical techniques have been embedded in a wide range of applications for the development of financial services in credit scoring and risk assessment \cite{sudjianto2010statistical,devi2018survey}. However, such models often struggle to represent complex financial patterns because they rely on fixed functions and statistical assumptions \cite{luo2017deep}. While they have some advantages such as transparency and interpretability, their performance tends to suffer when faced with the challenges presented by the vast amounts of data and intricate relationships in credit prediction tasks.

On the contrary, Deep Learning (DL) approaches have garnered significant attention across diverse domains, including the financial sector. This is due to their superior performance compared to traditional statistical and Machine Learning (ML) models \cite{teles2020machine}. 
In particular, DL has made great strides in several application areas, such as medical imaging \cite{10371865}, price forecasting \cite{lago2018forecasting} \cite{10.1007/978-3-030-05918-7_25}, and structural health monitoring \cite{data23} \cite{10.1007/978-3-031-37317-6_5} \cite{ncta22} \cite{delta22}, demonstrating its versatility in handling complex data patterns.

Besides developing classification strategies, a distinct approach to enhance the workflow is to focus on preprocessing. A common data preprocessing technique in the credit scoring field is Weight of Evidence (WoE) data encoding, as it enjoys several properties \cite{thomas2017credit}. First, being a target-encoding method, is able to capture nonlinear relationships between the features and the target variable. Second, it can handle missing values; which often afflict credit scoring datasets as borrowers may not provide all the required information when applying for a loan. WoE handles missing values by binning them separately. Finally, WoE coding reduces data dimensionality by scaling features (both numerical and categorical) into a single continuous variable. This can be particularly useful in statistic, ML and DL contexts, because models may have different intrinsic structures and may only be able to work with a specific data type \cite{l2017machine}.

The goal of this work is to combine different technologies and frameworks into an effective ML workflow to address the task of credit default prediction for the financial sector. Besides the data preprocessing via WoE coding, we introduce an ensemble strategy to build a more robust model; a hyperparameter optimization to maximize performance, and a loss function that focuses learning on hard-to-classify examples to overcome data imbalance problems.

To assess model performance and workflow strength, we present results obtained on known and publicly available benchmark datasets. These datasets provide a common reference point and enable meaningful comparisons between different models.

The paper is organized as follows. The material and methodology are covered in Section 2, while the experiment results and discussions are covered in Section 3. Finally, Section 4 draws conclusions and outlines avenues for future research.


\section{Materials and methodology}\label{sec:methodology}

\begin{figure*}[!b]
\centerline{\includegraphics[width=0.9\textwidth]{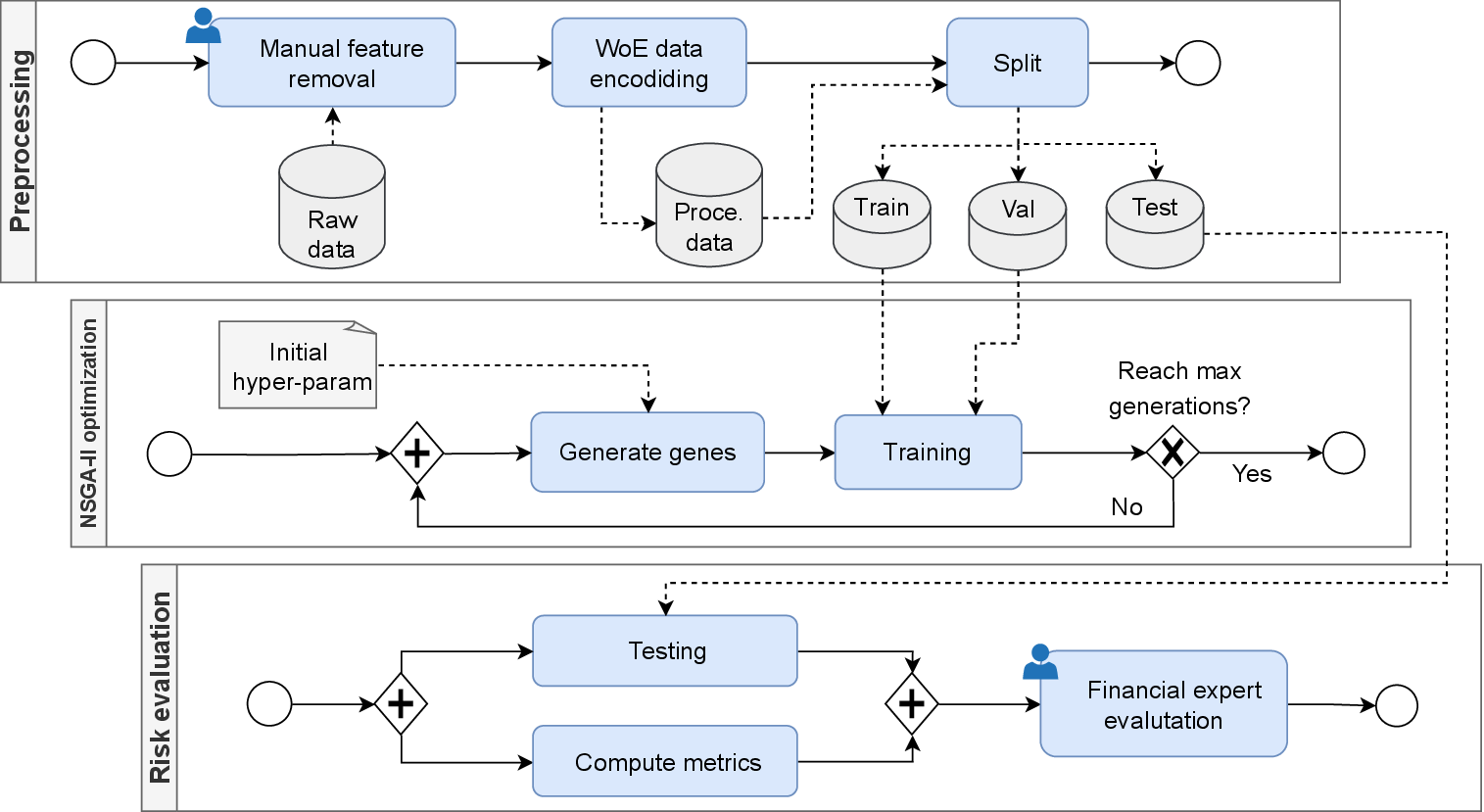}}
\caption{Workflow design of the proposed method.}
\label{fig:bpmn-plot}
\end{figure*}

The proposed ML workflow is shown in Figure \ref{fig:bpmn-plot} by means of a Business Process Model and Notation (BPMN) diagram. BPMN is a formal graphical notation that provides a visual representation of business processes and workflows, allowing for efficient interpretation and analysis of systems \cite{cimino2014interval}. BPMN was chosen due to its ability to visually represent complex processes in a standardized and easily understandable manner. 

The diagram provides a comprehensive overview of the ML workflow for credit scoring default prediction tasks. 
The first lane focuses on data preprocessing, where manual column removal and data encoding through Weight of Evidence (WOE) techniques are employed. The second lane is dedicated to model training and optimization, exploring various learning models described below. Finally, the third lane involves computing evaluation metrics, while also incorporating the expertise of a financial expert to assess the performance.

The second lane aims to solve a supervised machine learning problem where the goal is to predict whether a borrower is likely to default on a loan or not. Specifically, a binary classification model \cite{dastile2020statistical}, trained on a dataset of historical borrowers information with the final goal of finding a model $\psi_p : \mathbb{R}^n \Rightarrow \{-1, +1\}$ which maps a feature vector $x \in \mathbb{R}^n$ to an output class $y \in \{-1, +1\}$; where $x$ identifies the set of attributes describing a borrower, $y$ is the class label (non-default $-1$, default $+1$), and $p$ is the set of parameters describing the model $\psi$:
\begin{equation}\label{eq:bin_classifier}
\psi_p : x\Rightarrow y.
\end{equation}

To evaluate the classification performance of the above problem, the Area Under the Curve (AUC) metric is introduced:

\begin{equation}
AUC = \int_{0}^{1} ROC(u) \ du,
\end{equation}

\noindent
where $ROC(u)$ is the receiver operating characteristic (ROC) curve at threshold $u$, defined as the ratio between the true positive rate $TPR(u)$ and the false positive rate $FPR(u)$ both at threshold $u$.

Another popular metric for evaluating performance when dealing with unbalanced datasets is the F-score, computed as the average of the well-known $precision$ and $recall$ metrics.

The Brier score metric \cite{beque2017approaches} was used to measure the mean squared difference between the predicted probability and the actual outcome. Given a dataset $\mathcal{D}$, composed of $n$ samples, BS metric is shown in Equation \ref{eq:bs}.
\begin{equation}\label{eq:bs}
    BS = \frac{1}{n} \sum_{i = 1}^{n} \left(p_i - o_i\right)^2,
\end{equation}
where $p_i$ is the (default) probability predicted by the model and $o_i$ is the actual label.

Generally in the credit scoring literature, the cost of incorrectly classifying a good applicant as a defaulter (i.e., $c_0$, false positive) is not considered to be as important as the cost of misclassifying a default applicant as good (i.e., $c_1$, false negative). Indeed, when a bad borrower is misclassified as good, they are granted a loan they are unlikely to repay, which can lead to significant financial losses for the lender \cite{hand2009measuring}. The $c_0$ cost is equal to the return on investment (ROI) of the loan and we assume the ROI ($c_0$) to be constant for all loans, as is usually the case in consumer credit scoring \cite{VERBRAKEN2014505}. It is worth noting that the above argument assumes that there is no opportunity cost associated with not granting a loan to a good credit borrower. However, in reality, there may be some opportunity cost, as the borrower may take their business elsewhere if they are not granted a loan \cite{verbraken2014development}.

Under this premise, we introduce the Expected Maximum Profit (EMP) metric, since the metrics introduced previously consider only minimizing credit risk and not necessarily maximizing the profit of the lender. The EMP metric takes into account both the probability of insolvency and the profit associated with each loan decision \cite{oskarsdottir2019value}. 

To define the EMP metric we first introduce the average classification profit metric per borrower in Equation \ref{eq:avg_classification_profit}; it is determined based on the prior probabilities of defaulters $p_0$ and non-defaulters $p_1$, as well as the cumulative density functions of defaulters $F_0$ and non-defaulters $F_1$. Additionally, $b_0$ represents the profit gained from correctly identifying a defaulter, $c_1$ denotes the cost incurred from erroneously classifying a non-defaulter as a defaulter, while $c*$ refers to the cost associated with the action taken. Hence, EMP can be defined as shown in Equation \ref{eq:emp}:
\begin{equation}\label{eq:avg_classification_profit}
P(t;b_0,b_1,c*) = (b_0-c*)\pi_0 F_0 (t) - (c_1-c*)\pi_1 F_1 (t)
\end{equation}
\begin{equation}\label{eq:emp}
    EMP = \int_{b_0}\int_{c_1} P(T(\theta); b_0, c_1, c*) \cdot h(b_0, c_1)\ db_0\ cd_1
\end{equation}

where $\theta = \frac{c_1+c*}{b_0-c*}$ is the cost-benefit ratio, while $h(b_0, c_1)$ is the joint probability density function of the classification costs. Finally, the best cut-off value is $T$ as shown in Equation \ref{eq:cut-off}; and, the average cut-off-dependent classification profit is optimized to produce the highest profit.
\begin{equation}\label{eq:cut-off}
    T = \argmax_{\forall t} P(t;b_0,b_1,c*)
\end{equation}

\subsection{Learning models}
According to \cite{dastile2020statistical}, in this section, we introduce three categories of learning models: statistical models, machine learning, and deep learning.

Logistic regression (LR) is a popular statistical model in binary classification defined by the formulas $P(y=1 | x) = \frac{1}{1+exp{(-(\alpha_0 + \alpha^T x))}}$ and $P(y=-1 | x) = 1 - P(y=1 | x)$; where $P(y=1 | x)$ and $P(y=-1 | x)$ are the probabilities of classifying the observation $x$ as a good or bad borrower, respectively. Once the model parameters $\alpha_0$ and $\alpha$ are trained, the decision rule to classify an input feature vector $x$ as the output value $y$ is 
\begin{equation}\label{eq:log_regression_rule}
y =
\begin{cases}
+1 & \text{when } exp{(\alpha_0 + \alpha^T x)} < 1\\
-1 & \text{otherwise.}
\end{cases}
\end{equation}

Another category of models introduced is the ML ones. A Classification Tree (CT) is a popular algorithm used as a classifier in ML. It is a flowchart-like structure, where each internal node represents a feature, each branch represents a decision rule, and each leaf node represents the classification. The algorithm works by recursively partitioning the dataset, based on the feature that best splits the data at each node, until a stopping criterion is reached. 

The last model category introduced is DL, which through neural networks outperformed in several areas compared to traditional models. This is due to DL's ability to learn hierarchical representations and complex patterns of input data. 

Each learning model can be enhanced with the ensemble technique. This approach combines the predictions of multiple models to improve the overall classification performance. Specifically, a weight-based voting strategy is implemented to combine the predictions. The decision function of the ensemble models can be expressed as:
\begin{equation}\label{eq:ensemble}
    y = \argmax \sum_{i=1}^{n} a_i \cdot w_i
\end{equation}
where $a_i$ is the predicted class probability by the $i$-th individual model, and $w_i$ is the weight assigned to the $i$-th model.

In the case of ensemble of different CT, a model called Random Forrest (RF) is obtained; while in the case of the DL ensemble, it is referred to as Ensemble Multi-Layer Perceptron (EMLP).

\subsection{Data encoding}
The Weight of Evidence (WoE) encoding was used as a data encoding method to preprocess the datasets \cite{raymaekers2022weight}. The WoE value of each categorical variable is computed as:
\begin{equation}
WoE_i = \ln\left(\frac{P_{i,0}}{P_{i,1}}\right)
\end{equation}
where $WoE_i$ is the WoE value for category $i$, $P_{i,1}$ is the probability of a borrower defaulting on a loan within category $i$, and $P_{i,0}$ is the probability of a borrower not defaulting on a loan.

WoE encoding can also be applied to numerical variables, by first discretizing them through binning process. It does not embed a binning strategy, hence it must be explicitly defined and integrated within the data encoding. Several binning techniques have been devised, such as equal-width or equal-size, however, not all of them guarantee the necessary conditions for good binning in credit scoring \cite{zeng2014necessary}: 
\begin{itemize}
    \item missing values are binned separately,
    \item a minimum of 5\% of the observations per bin, 
    \item for either good or bad, no bins have 0 accounts.
\end{itemize}

\noindent
In the proposed workflow, we integrated the optimal binning method proposed by Palencia; his implementation is publicly available at \cite{optbinning-repo}. The optimal binning algorithm involves two steps: A prebinning procedure generating an initial granular discretization and further fine-tuning to satisfy the enforced constraints.

The implementation proposed by Palencia is based on the formulation of a mathematical optimization problem solvable by mixed-integer programming in \cite{Navas-Palencia2020OptBinning}. The formulation was provided for a binary, continuous, and multi-class target type and guaranteed an optimal solution for a given set of input parameters. Moreover, the mathematical formulation of the problem is convex, resulting that there is only one optimal solution that can be obtained efficiently by standard optimization methods.

\subsection{Hyperparameter optimization}
Non-dominated Sorting Genetic Algorithm II (NSGA-II) was introduced in the workflow to perform the hyperparameter optimization of credit scoring models \cite{verma2021comprehensive}. NSGA-II is a well-known multi-objective optimization algorithm widely used in various domains. In the workflow, we used NSGA-II to optimize the hyperparameters of the models, by considering two distinct objective functions: the Area Under the Receiver Operating Characteristic curve (AUC) as a classification metric, and the Expected Maximum Profit (EMP) as a financial metric. By incorporating EMP, we aim to optimize the credit scoring models not only for classification accuracy but also for their financial impact. The proposed approach enables us to find a set of non-dominated solutions that provide the best trade-off between AUC and EMP and allows us to select the best model for a particular financial institution based on their specific requirements.

\subsection{Focal loss}
It has been shown that class imbalance impedes classification. However, we refrain from balancing classes for two reasons. First, our objective is to examine relative performance differences across different classifiers. If class imbalance hurts all classifiers in the same way, it would affect the absolute level of observed performance but not the relative performance differences among classifiers. Second, if some classifiers are particularly robust toward class imbalance, then such a trait is a relevant indicator of the classifier's merit. Equation \ref{eq:rate-def} presents the $rate_{def}$ indicator used to evaluate the dataset unbalance.
\begin{equation} \label{eq:rate-def}
rate_{def} = \frac{Default\ cases}{Total\ cases}
\end{equation}

To mitigate the problem, a loss function called $focal loss$ \cite{mukhoti2020calibrating} was used; Equation \ref{eq:focal-loss} shows its formulation. 

Focal loss is a modification of the cross-entropy loss function, which assigns a higher weight to hard examples that are misclassified. The focal loss also introduces the focusing parameter, which tunes the emphasis degree on misclassified samples. 
\begin{equation} \label{eq:focal-loss}
FL(p_t) = -\alpha_t (1-p_t)^\gamma ln(p_t)
\end{equation}
where $p_t$ is the predicted probability of the true class, $ \alpha_t \in [0,1]$ is a weighting factor for class $t$ and $\gamma$ is the focusing parameter.


\section{Experiments and results}\label{sec:experiments}

The described experiments were performed in Python programming language on a Jupyter Lab server running Arch Linux operating system. Hardware resources used included AMD Ryzen 9 5950x CPU, Nvidia RTX A5000 GPU and 128 GiB of RAM. To ensure reproducibility and transparency, we publicly released the code and results of the experiments on GitHub. 

Four datasets well-known in the literature and publicly available were used to implement and test the proposed methodology. 
Table \ref{tab:dataset-information} presents the datasets indicating the amount of samples and the $rate_{def}$.

\begin{table}[h]
\centering
\caption{Dataset details}
\begin{tabular}{|lcc|}
\hline
\textbf{Name} & \textbf{Cases}  & \textbf{rate$_{\mathbf{def}}$} \\
\hline
GermanCreditData-GER & 1000 & 0.3 \\
HomeEquityLoans-HEL & 5960 & 0.19 \\
HomeEquityCreditLine-HECL & 10460 & 0.52 \\
PolishBankruptcyData-PBD & 43405 & 0.04 \\
\hline
\end{tabular}
\label{tab:dataset-information}
\end{table}

The GER and PBD datasets are popular credit scoring data accessible through the UCI Machine Learning repository\footnote{https:\/\/archive.ics.uci.edu\/}. The HEL dataset was released publicly in 2020 with \cite{do2020liquidity}. The HELC dataset was provided by Fair Isaac Corporation (FICO) as part of the  Explainable Machine Learning challenge\footnote{https:\/\/community.fico.com\/s\/explainable-machine-learning-challenge}.

To ensure that good estimates of the performance of each classifier are obtained, Optuna \cite{akiba2019optuna}, an open source hyperparameter optimization software framework, was used. Optuna enables efficient hyperparameter optimization by adopting state-of-the-art algorithms for sampling hyperparameters and pruning efficiently unpromising trials. The provided NSGA-II implementation with default parameters was used to continually narrow down the search space leading to better objective values.

\begin{figure}[h]
\begin{center}
    
   \includegraphics[width=.48\textwidth]{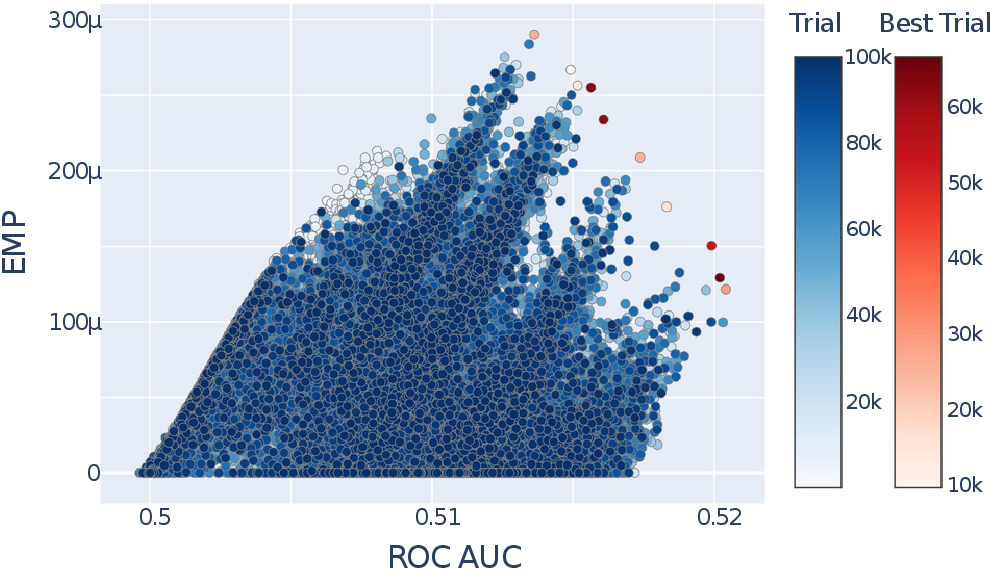}
   
   \caption{Scatter plot of the random forrest hyperparameter optimization process}
   \label{fig:pareto}
\end{center}
\end{figure}

Figure \ref{fig:pareto} illustrates an example of hyperparameter optimization processes and highlights the pareto front, represented by the red points in the scatter plot. The pareto front is composed of the non-dominated solutions that refer to the best sets of hyperparameters, capturing the trade-off between EMP and AUC performance metrics \cite{hua2021survey}. The models whose results are shown in Tables \ref{tab:GER_metrics}, \ref{tab:HEL_metrics}, \ref{tab:HECL_metrics} and \ref{tab:PBD_metrics} were manually chosen from those on the pareto front by observing the values of the performance metrics.

We can see how the DL models outperformed the statistical and ML models for each dataset; in fact, the best results are consistently found in the last rows of the tables for the MLP and EMLP models. In addition, the ensemble models introduce an enhancement over the corresponding non-ensemble models.

\begin{table}[h!]
\begin{center}
\caption{Performance metrics on GER dataset.}\label{tab:GER_metrics}
\begin{tabular}{|*{5}{>{\centering\arraybackslash}p{1.0cm}}|}
\hline
Model &  AUC & F1 & BS & EMP\\
\hline 
 LR & .800 & .627 & .255 & .051\\
 CT & .701 & .546 & .341 & .041  \\
 RF & .792 & .558 & .236 & .037  \\
 MLP & .799 & .616 & .273 & .050  \\
 EMLP & .801 & .632 & .249 & .053 \\
  \hline
\end{tabular}
\end{center}
\end{table}

\begin{table}[h!]
\begin{center}
\caption{Performance metrics on HEL dataset.}\label{tab:HEL_metrics}
\begin{tabular}{|*{5}{>{\centering\arraybackslash}p{1.0cm}}|}
\hline
Model &  AUC & F1 & BS & EMP\\
\hline 
 LR & .869 & .580 & .151 & .017 \\
 CT & .820 & .671 & .152 & .025 \\
 RF & .940 & .693 & .114 & .023 \\
 MLP & .864 & .604 & .210 & .022 \\
 EMLP & .866 & .636 & .136 & .024 \\
  \hline
\end{tabular}
\end{center}
\end{table}

\begin{table}[h!]
\begin{center}
\caption{Performance metrics on HECL dataset.}\label{tab:HECL_metrics}
\begin{tabular}{|*{5}{>{\centering\arraybackslash}p{1.0cm}}|}
\hline
Model &  AUC & F1 & BS & EMP\\
\hline 
 LR & .801 & .610 & .251 & .054 \\
 CT & .812 & .631 & .242 & .060 \\
 RF & .863 & .703 & .214 & .063 \\
 MLP & .892 & .717 & .198 & .068 \\
 EMLP & .906 & .748 & .136 & .070 \\
  \hline
\end{tabular}
\end{center}
\end{table}

\begin{table}[h!]
\begin{center}
\caption{Performance metrics on PBD dataset.}\label{tab:PBD_metrics}
\begin{tabular}{|*{5}{>{\centering\arraybackslash}p{1.cm}}|}
\hline
Model &  AUC & F1 & BS & EMP\\
\hline 
 LR & .781 & .516 & .359 & .051 \\
 CT & .793 & .538 & .342 & .059 \\
 RF & .824 & .609 & .317 & .060 \\
 MLP & .841 & .612 & .296 & .062 \\
 EMLP & .883 & .648 & .233 & .069 \\
  \hline
\end{tabular}
\end{center}
\end{table}


\section{Conclusion}\label{sec:conclusion}

In this paper, we proposed a novel ML workflow for assessing the risk evaluation in the credit scoring context that combines WoE-based preprocessing, ensemble strategies of different learning models, and NSGA-II hyperparameter optimization. 

The proposed workflow has been tested on different public datasets, and we have presented benchmarks. The experiments indicate the methodology succeeds in effectively combining the strengths of the different technologies and frameworks that constitute the workflow to improve the robustness and reliability of the risk assessment support tools in the financial industry.

Future work could explore the applicability of our approach in real-world scenarios by integrating the classification models into enterprise software systems, thereby enhancing usability for bank employees and financial consultants. This integration has the potential to streamline and optimize financial processes, providing a practical solution for the challenges faced in the banking and financial consulting domains.
In addition, the applicability of this approach can be extended to corporate credit scoring, beyond the customer.


\section*{\uppercase{Acknowledgements}}

Work partially supported by: (i) the University of Pisa, in the framework of the PRA 2022 101 project “Decision Support Systems for territorial networks for managing ecosystem services”; (ii) the European Commission under the NextGenerationEU program, Partenariato Esteso PNRR PE1 - "FAIR - Future Artificial Intelligence Research" - Spoke 1 "Human-centered AI"; (iii) the Italian Ministry of Education and Research (MIUR) in the framework of the FoReLab project (Departments of Excellence) and of the "Reasoning" project, PRIN 2020 LS Programme, Project number 2493 04-11-2021.

\bibliographystyle{apalike}
{\small
\bibliography{bibliography}}

\end{document}